\documentstyle[12pt,epsf]{article}

\newcommand{\be}{\begin{equation}}
\newcommand{\ee}{\end{equation}}
\newcommand{\ben}{\begin{eqnarray}\displaystyle}
\newcommand{\een}{\end{eqnarray}}
\newcommand{\refb}[1]{(\ref{#1})} 
\newcommand{\p}{\partial}
\newcommand{\sectiono}[1]{\section{#1}\setcounter{equation}{0}}

\newcommand{\rr}{R}
\newcommand{\II}{{\cal I}}
\newcommand{\CC}{{\cal C}}
\newcommand{\SS}{{\cal S}}

\newcommand{\HH}{{\cal H}}

\newcommand{\NN}{{\cal N}}

\newcommand{\RR}{{\cal R}}

\newcommand{\QQ}{{\cal Q}}
  
\newcommand{\bsliv}{\langle\Xi |}

\newcommand{\wt}{\widetilde}
\newcommand{\wh}{\widehat}

\newcommand{\bmu}{{\bar\mu}}
\newcommand{\bnu}{{\bar\nu}}
\newcommand{\xl}{{x^L}}
\newcommand{\xr}{{ x^R}}

\newcommand{\vp}{\varphi}

\begin{document}

{}~
\hfill\vbox{\hbox{hep-th/0106010}\hbox{CTP-MIT-3151}
\hbox{PUPT-1992} \hbox{NSF-ITP-01-53}
}\break

\vskip .6cm

\centerline{\large \bf 
Vacuum String Field Theory} 

\vspace*{4.0ex}

\centerline{\large \rm Leonardo Rastelli$^a$, 
Ashoke Sen$^b$ and Barton
Zwiebach$^c$}

\vspace*{4.0ex}

\centerline{\large \it ~$^a$Department of Physics }

\centerline{\large \it Princeton University, Princeton, NJ 08540,
USA}

\centerline{E-mail:
        rastelli@feynman.princeton.edu}

\vspace*{2ex}

\centerline{\large \it ~$^b$Harish-Chandra Research
Institute}

\centerline{\large \it  Chhatnag Road, Jhusi,
Allahabad 211019, INDIA}

\centerline {and}
\centerline{\large \it Institute for Theoretical Physics}

\centerline{\large \it 
University of California, Santa Barbara, CA 93106, USA
}

\centerline{E-mail: asen@thwgs.cern.ch, sen@mri.ernet.in}

\vspace*{2ex}

\centerline{\large \it $^c$Center for Theoretical Physics}

\centerline{\large \it
Massachussetts Institute of Technology,}

\centerline{\large \it Cambridge,
MA 02139, USA}

\centerline{E-mail: zwiebach@mitlns.mit.edu}

\vspace*{5.0ex}

\centerline{\bf Abstract}
\bigskip

This is a brief review of vacuum string field 
theory, a new approach to open string field
theory based on the stable vacuum of the tachyon.
We discuss the sliver state 
explaining  its role as a projector in the space of half-string 
functionals. We review the construction of D-brane solutions 
in vacuum string field theory, both in the algebraic 
approach and in the more general geometrical approach 
that emphasizes the role of boundary CFT. 
{\it (To appear in the Proceedings of Strings 2001, Mumbai, India)}.

\vfill \eject

\baselineskip=16pt

\tableofcontents

\newpage

\sectiono{Introduction and summary}  \label{s1}

Much of the work  done to understand 
various conjectures about tachyon
condensation  in bosonic string
theory \cite{conj}  has used
cubic open string field
theory \cite{WITTENBSFT} (OSFT).
Although the results are very 
impressive, they ultimately rely 
on numerical study of the solutions
of the equations of motion using the level truncation  
scheme~\cite{KS,9912249}.\footnote{For early
attempts at understanding the open string  tachyon, 
see refs.\cite{EARLY}.
For field theory models of tachyon condensation, see 
refs.\cite{FIELD1,FIELD2}. Studies  using 
renormalization  group  have been carried 
out in refs.\cite{RG}. For boundary string field theory
studies of open string tachyons, see refs.~\cite{BOLD,BNEW,BLOOP}.} 
In  a series of papers~\cite{0012251,0102112,0105058,0105168} we 
proposed an analytic
approach to  tachyon condensation  
by introducing a new approach to open string
field theory $-$ vacuum string field theory 
(VSFT).\footnote{Since  
the time of Strings 2001, the subject has developed rapidly.
Instead of just summarizing the talks given at the conference, we
shall try to give a summary of the current state of knowledge in
this subject.} 
This theory uses the open string tachyon vacuum
to formulate the     dynamics. 
Among all possible open string backgrounds the tachyon vacuum is
particularly natural given its physically expected uniqueness as the
endpoint of all processes of tachyon condensation. 
 As opposed to the conventional OSFT, where the
kinetic operator is the BRST operator $Q_B$, in VSFT the kinetic
operator
$\QQ$ is non-dynamical and is built solely out of worldsheet ghost
fields.\footnote{A subset
of this class of actions was discussed previously 
in ref.~\cite{HORO}.}
In this class of 
actions gauge invariance is naturally achieved, and
the absence of physical open string states around the vacuum is
manifest. Work related to our 
ref.\cite{0105058} has also been
carried out by Gross and Taylor~\cite{0105059}.
Additional recent work is found in \cite{0105175,0105184}.
Related questions have been addressed in the boundary string
field theory approach in ref.\cite{0105245}.  

It is now clear that VSFT is structurally
much simpler than conventional OSFT.
Indeed, it is possible to
construct analytically classical solutions
representing arbitrary D-$p$ branes, with correct
ratios of tensions, thereby providing a non-trivial check on 
the correctness of  our proposal. 
The key ansatz that makes this analysis possible is 
that the string field solution 
representing a D-brane factorizes into a ghost part
$\Psi_g$ and a matter part $\Psi_m$, with $\Psi_g$ 
the same for
all D-branes, and $\Psi_m$ 
different for different D-branes. As of this writing,
however,  the 
universal ghost string field $\Psi_g$ is
still unknown. Moreover, the specific choice of $\QQ$
in the VSFT action is also unclear. These are probably the central
open questions in this formulation of string field theory.

The matter part of the string field satisfies a very simple
equation: it squares to itself under $*$-multiplication.
Two points of view have been useful
for solving 
this equation.  In the geometric method,
the $*$-product is defined by
the gluing of Riemann surfaces.
In the algebraic method,
one relies on the operator representation of the
$*$-product using flat space oscillator modes.
In both approaches a key role is played
by the sliver state, a solution of the
matter string field equations which can alternatively
be viewed geometrically as the surface state associated
with a specific one-punctured disk, or algebraically
as a squeezed state, {\it i.e.}  
the exponential of an oscillator bilinear acting on the vacuum.  
Solutions representing various
(multiple) D-branes in vacuum string field theory
are obtained as  (superpositions of) 
various deformations of the sliver state.

The geometric approach turns out to be very flexible
in that it allows the construction of (multiple) 
D-brane solutions corresponding to arbitrary boundary CFT's, 
in any spacetime background. The correct ratios of tensions are 
obtained manifestly because the norm of the sliver
solution is naturally related to the disk partition
function of the appropriate boundary CFT.
The geometric
picture of the sliver also 
indicates that it is a state originating from
a Riemann surface where the left-half and the right-half 
of the  open 
string are ``as
far as they can be'' from each other. 
Indeed, the sliver functional 
factors into the product of 
functionals
of the left-half  and  the 
right half of the open string,
and hence can be thought as a rank-one projector
in a space of half-string functionals. 

The algebraic approach, while more tied to
the choice of a flat-space background, provides
very explicit expressions for the string fields.
More crucially, in the algebraic approach we can 
take direct advantage of the insight that D-brane
solutions are projectors onto 
the half-string state space. The intuitive left/right
splitting picture 
provided by the functional representation
can be turned into a completely algebraic
procedure to obtain multiple D-brane solutions
of various dimensions, situated at various 
positions.

\sectiono{Vacuum string field theory} 
\label{sr1}

In this section we shall briefly review
the basic setup introduced in refs.\cite{0012251,0102112}.

\subsection{The gauge invariant action} 

In order to write concretely the string field theory
action we need
to use the state space $\HH$ of some speficic 
matter-ghost boundary conformal field theory (BCFT).
For this, we shall consider a 
general space-time background described by some arbitrary bulk CFT, 
and we pick some fixed D-brane  associated
with a specific BCFT. We shall call this BCFT$_0$ and $\HH$
its state space\footnote{In \cite{0012251, 0102112}
BCFT$_0$ was taken to be the D25 brane in flat background.}.
The string
field  $\Psi$ is a state of ghost number one  
in $\HH$  and the
string field action is given by: 
\be \label{eo1}
\SS (\Psi) \equiv \,-\, {1\over g_0^2}\,\,\bigg[\, 
{1\over 2}
\langle
\,\Psi \, ,
 \, \QQ\, \Psi
\rangle + {1\over 3}\langle \,\Psi \, , \, \Psi *
\Psi \rangle \bigg] \,,
\ee
where $g_0$ is the open
string coupling constant, $\QQ$ is an operator of 
ghost number one, 
$\langle \, ,  \, \rangle$
denotes the BPZ inner product,
and $*$ denotes the usual $*$-product of the string
fields~\cite{WITTENBSFT}. 
$\QQ$ satisfies the requirements:
\ben \label{eFINp}
&& \QQ^2 = 0, \nonumber \\
&& \QQ (A * B) = (\QQ A) * B + (-1)^{A} A * (\QQ B)\, , \\
&& \langle \, \QQ A , B \,\rangle = - (-)^A \langle A , \QQ B \rangle
\,. \nonumber
\een 
The action \refb{eo1} is then invariant under the 
gauge transformation:
\be \label{egtrs}
\delta\Psi = \QQ\Lambda + \Psi * \Lambda - \Lambda * \Psi \, ,
\ee
for any ghost number zero state $\Lambda$ in $\HH$. 
Besides obeying the conditions (\ref{eFINp}) $-$
the generic algebraic constraints that
guarantee gauge invariance,
the operator $\QQ$ is required to satisfy 
two physical requirements which are expected
for the tachyon vacuum:

\begin{itemize}

\item  The operator $\QQ$ must have vanishing cohomology.

\item The operator $\QQ$ must be universal, 
namely, it must be possible to
express $\QQ$ 
without reference to the reference 
boundary conformal field theory BCFT$_0$. 

\end{itemize}

We can satisfy these requirements by letting
$\QQ$ be constructed purely from ghost operators. 
In particular any linear combination of 
the ghost number one operators
\be
\label{cn}
\CC_n \equiv  c_n +  (-)^n \, c_{-n}  \,, \quad n=0,1,2,\cdots
\ee
satisfies the required properties. As discussed
in \cite{0012251} many, but not all,
of these operators are related by homogeneous
field redefinitions of the form 
$\Psi \to \exp(\sum v_n K_n) \Psi$,
where $K_n = L_n - (-1)^n L_{-n}$ generate 
the
reparametrization invariances of the cubic vertex.

Broadly speaking, there are two classes of $\QQ$ operators, $-$
ones which annihilate the identity string field $|\II\rangle$, and
ones which don't. The simplest operator, $\QQ=c_0$, does not
annihilate $|\II\rangle$, but $\QQ= c_0+1/2(c_2 + c_{-2})$ does.
Both kinds of operators lead to gauge invariant actions. While
$\QQ$'s which do annihilate the identity seem somewhat more
regular, $\QQ$'s which do not annihilate the identity may be
needed to obtain a consistent
formulation of the ghost sector.

\subsection{Factorization ansatz for D-brane solutions}

In ref.~\cite{0102112} we postulated that all D-$p$-brane
solutions of VSFT have the factorized 
form:\footnote{We thank W. Taylor 
for pointing out this possibility.}
\be \label{eo3}
\Psi = \Psi_g \otimes \Psi_m\, ,
\ee
where $\Psi_g$
denotes a state obtained by acting with the ghost
oscillators on the SL(2,R) invariant vacuum of the ghost 
sector of BCFT$_0$, and  
$\Psi_m$  is a
state obtained by acting with matter operators on the SL(2,R)
invariant
vacuum of the matter sector of BCFT$_0$.  
Let us denote by
$*^g$ and $*^m$ the star product in the ghost and matter sector
respectively. The equations of motion 
\be \label{eo2}
\QQ \Psi + \Psi * \Psi = 0\, 
\ee
factorize as
\be \label{eo4}
\QQ \Psi_g = - \Psi_g *^g \Psi_g \,,
\ee
and
\be \label{eo5}
\Psi_m = \Psi_m *^m \Psi_m\, .
\ee
We further assumed that the ghost part $\Psi_g$ is universal for all
D-$p$-brane solutions. Under this assumption the ratio of energies
associated with two different D-brane solutions, with matter parts
$\Psi_m'$ and $\Psi_m$ respectively, is given by:
\be \label{eo7}
{\langle \Psi_m' | \Psi_m'\rangle_m \over \langle \Psi_m |
\Psi_m\rangle_m} \, ,
\ee
with $\langle \cdot| \cdot\rangle_m$ denoting BPZ inner product in
the matter BCFT. Thus the ghost part drops out of this calculation.

\sectiono{Viewpoints for the sliver}

The sliver state $| \Xi \rangle$ is a ghost number zero
state satisfying the equation $\Xi * \Xi = \Xi$.
Furthermore it can be written in a factorized form:
$\Xi=\Xi_g\otimes \Xi_m$. Thus if we normalize $\Xi_g$ so
that it squares to itself under the $*$-product, then the
matter part $\Xi_m$ of the sliver provides a
solution of the string
field equations (\ref{eo5}).
This describes a configuration corresponding to a single D-brane.
We first review the universal geometric definition of
the sliver as a surface state \cite{0006240, 0105168}, 
and then its description in oscillator language 
(in a flat background) as a squeezed state 
\cite{0008252, 0102112}.

\subsection{The sliver as a surface state}

\label{sursl}

The sliver is a ghost number zero state that has a 
universal 
definition. It is a {\it surface state},
which means that for any given BCFT 
it can be defined as the bra $\langle \Xi|$
associated to a particular Riemann surface $\Sigma$.  
The surface in question
is a disk $D$ with one puncture $P$ at the boundary. 
Moreover, there
is a local coordinate at this puncture.

\newcommand{\lo}{\xi}
\newcommand{\gl}{z}
\newcommand{\gf}{\wh w}

\subsubsection{Surface states in boundary CFT} 
\label{sursl1}

\begin{figure}[!ht]
\leavevmode
\begin{center}
\epsfxsize = 15 cm \epsfbox{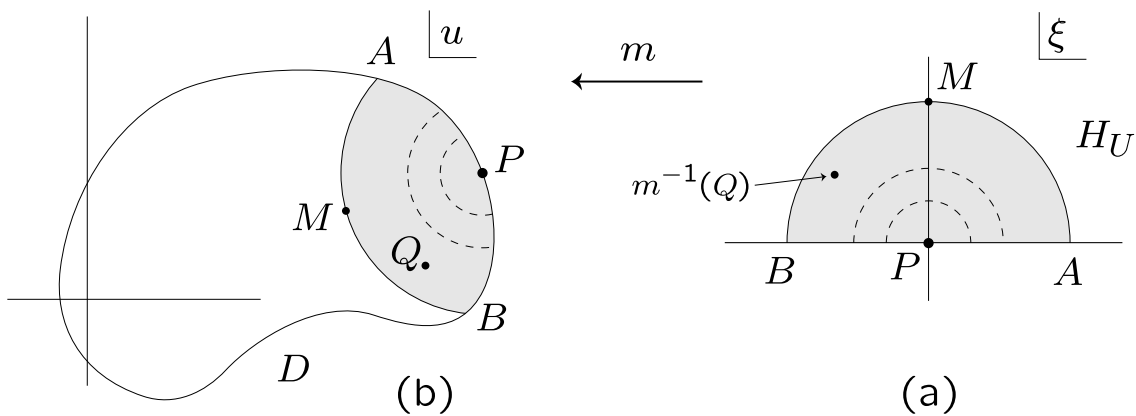 }
\end{center}
\caption[]{\small A punctured disk $D$ with a local coordinate
around the puncture $P$. The coordinate is defined via a map
from the half-disk $H_U$ to $D$. The arcs $AM$ and $MB$ in 
$D$ represent the left half and the right half of the open string
respectively. } \label{f1}
\end{figure}

Let us first discuss general surface states 
associated with a disk with one puncture. 
A local coordinate at a puncture (see Fig.~\ref{f1}) is obtained from
an analytic map $m$ taking a canonical half-disk $H_U$
defined as 
\be
\label{halfdisk}
H_U: \,\,\{ |\lo|\leq 1, \Im (\lo)\geq 0\}\,,
\ee
{\it into} $D$, where $\lo=0$ maps to the puncture $P$, and
the image of the real segment
$\{|\lo|\leq 1, \Im (\lo) =0\}$ lies on the boundary of $D$. 
The coordinate $\lo$ of the half disk is
called the {\it local coordinate}. For any point $Q\in D$ in the
image of the map, 
$\lo( m^{-1}(Q))$ 
is the local coordinate of the point.
Using any {\it global  
coordinate} $u$ on the disk $D$, 
the map $m$ can be
described by some analytic function $s$: 
\be u = s(\lo)\,, \quad  u (P)= s(0) \,.
\ee
Given this geometrical data, and a BCFT with state space
${\cal H}$, the state 
$\langle\Sigma|\in {\cal H}^* $
associated to the surface $\Sigma$ is defined as follows.
For any local operator $\phi (\lo)$, with associated state
$|\phi\rangle = \lim_{\lo\to 0} \phi(\lo) |0\rangle$ we set
\be \label{esdef}
\langle\Sigma| \phi \rangle = \langle  s
\circ \phi (0) \rangle_{D} \,,
\ee
where $\langle ~\rangle_{D}$ 
corresponds to correlation function on $D$
and $s\circ \phi (0)$ 
denotes the conformal
transform of the operator by the map $s(\xi)$, {\it i.e.} the operator
$\phi(\lo=0)$
expressed using the appropriate conformal map in terms of
$\phi(s(0))$. 
For a
primary of dimension
$h$, $s\circ \phi (0) = \phi(s(0))
(s'(0))^h$. The right
hand side of eq.\refb{esdef} can be interpreted as
the one point
function on $D$ of the local operator $\phi$ inserted at $P$ using the
local coordinate
$\lo$ defined there. We also call, with a small abuse of notation,
$|\Sigma\rangle \in \HH$ a surface state; this is simply
the BPZ conjugate 
of $\langle \Sigma|$.  
While computations of correlation functions
involving states in $\HH$  requires that the map $s$ be defined
only locally around the puncture $P$, 
more general constructions, such as
the gluing of surfaces, an essential 
tool in the operator formulation of
CFT, requires that the full map of the half disk $H_U$ into
the disk $D$  be well
defined.

At an intuitive level
$\langle\Sigma|$ can  
be given the following functional integral representation. 
Consider the
path integral over the basic elementary fields of the two dimensional
conformal field theory, $-$ collectively denoted as $\vp$, $-$ on the
disk $D$ minus the local coordinate patch, with some fixed boundary
condition $\vp=\vp_0(\sigma)$ on the boundary $AMB$ of the local
coordinate patch, and the open string 
boundary condition corresponding to
the BCFT under study on the rest of the boundary of this region. The
parameter $\sigma$ is the coordinate labeling the open string along
$AMB$, defined  
through $\xi=e^{i\sigma}$. 
The result of 
this path integral
will be 
a functional of the boundary value $\vp_0(\sigma)$. We
identify this as the
wave-functional of the state $\langle\Sigma|$. (For describing the
wave-functional of $|\Sigma\rangle$ we need to make a
$\sigma\to(\pi-\sigma)$ transformation.) 
On the other hand the wave-functional
of the state $|\phi\rangle$ can be
obtained by performing the path integral over $\vp$ on the unit
half-disk in the $\xi$ coordinate system, with the boundary condition
$\vp=\vp_0(\sigma)$ on the
semicircle, open
string boundary condition corresponding to the BCFT
on real axis, and a vertex operator $\phi(0)$ inserted at the origin.
We can now compute
$\langle\Sigma|\phi\rangle$ for any state $|\phi\rangle$ in $\HH$ by
multiplying the two wave-functionals and integrating over
the argument 
$\vp_0(\sigma)$. The net result is a path integration
over $\vp$ on the full disk $D$, with the boundary condition
corresponding to BCFT over the full boundary and a vertex operator
$\phi$ inserted at the puncture $P$ {\it using the $\xi$ coordinate
system.} This is precisely eq.\refb{esdef}.

\subsubsection{The various pictures of the sliver}

We are now ready to define the sliver surface state.
Ref.\cite{0105168} describes several canonical
presentations of the sliver related
by conformal tranformations. Here we shall review
only three of them.

\newcommand{\ff}{\check f}

We begin by giving 
the description in which the disk $D$ is represented
as the unit disk $D_0: |w|\leq 1$
in a $w$-plane. The puncture will be located at $w=1$.
We define for any positive real number 
$n >  0$   
\be
\label{ih}
w_n = \ff_n (\xi) \equiv (h(\lo))^{2/n} =  
\Big({1+ i\lo\over 1-i\lo}\Big)^{2/n}\,, 
\ee
which for later purposes we also write as
\be
\label{ef1}
w_n = \exp \Bigl( \, i \,{4\over n} \, \tan^{-1} (\lo) \Bigr) \,.
\ee
The map $h(\lo)$ takes the canonical half disk into
a unit half disk in the $h(\xi)$-plane, 
lying in the region
$\Re(h(\xi))\ge 0$, $|h(\xi)|\le 1$, 
with the puncture at $h(0)=1$ 
on the
curved side of the half-disk. 
Moreover the
string midpoint $M$ at $\lo=i$  is mapped to $h(i)=0$. The map
$w_n = (h(\lo))^{2/n}$
makes the image of the canonical
half-disk into a wedge with the angle at $w_n=0$ equal to $2\pi/n$.
For any fixed $n$ we call the $\langle n|$ the
resulting surface state. Thus we have
\be \label{edefn}
\langle n| \phi\rangle 
\equiv \langle \ff_n\circ \phi(0)\rangle_{D_0}\,
\qquad \forall \phi .
\ee

\begin{figure}[!ht]
\leavevmode
\begin{center}
\epsfxsize = 15 cm \epsfbox{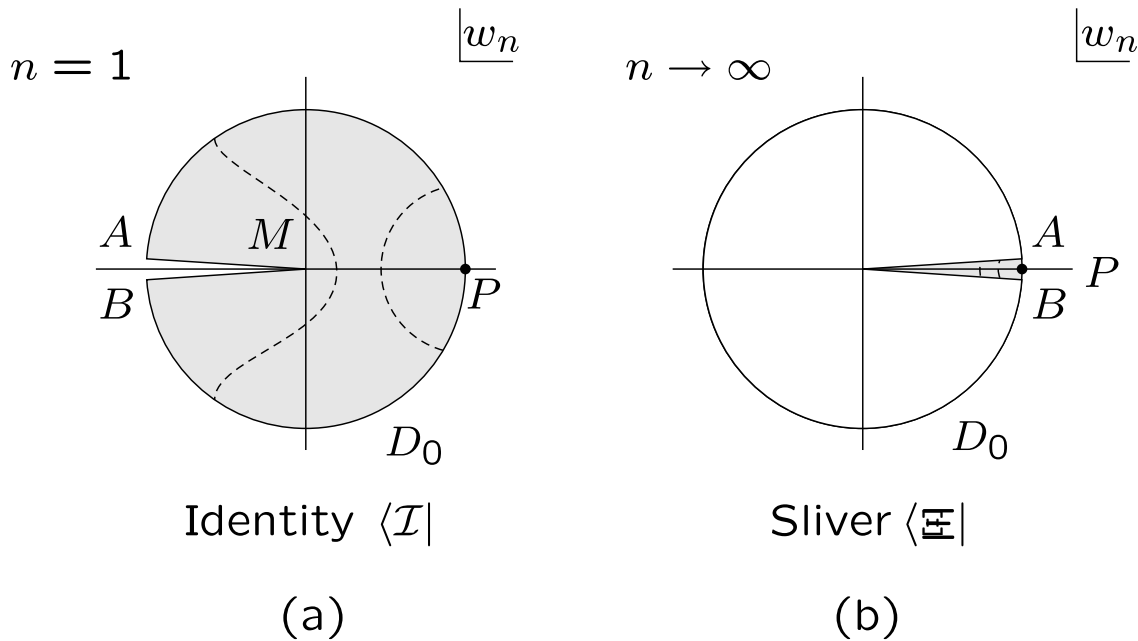 }
\end{center}
\caption[]{\small (a) The surface state corresponding to
the identity string field $\langle\II|$. Here the image
of $H_U$ covers the full disk, except for a cut in the
negative real axis. (b) The surfaces state corresponding
to the sliver $\langle \Xi|$. Here the image of $H_U$ covers
an infinitesimally thin sliver around the positive real axis. 
} \label{f4}
\end{figure}

The state obtained when $n=1$ is the
identity state (see Figure \ref{f4}-a). For this state the local
coordinate patch  
in the $w_n$ plane 
covers 
the full unit disk $D_0$ with a
cut on the
negative real axis. The left-half and the
right-half of the string coincide 
along this cut. The state $n=2$ is the vacuum state. In this case the
image of $H_U$ covers the right half of the full unit
disk $D_0$ in the $w_n$ plane. 
In the $n\to \infty$ limit, the image of $H_U$ 
in the $w_n$ coordinate is  a `thin sliver' of the  disk $D_0$
(Figure \ref{f4}-b). 
It was seen in \cite{0006240} and explained in
detail in \cite{0105168} that the limit $n\to \infty$ of
$\langle n|$ gives
rise to a well-defined state. The key is to use SL(2,R)
invarainces to resolve the apparent singularity in the
local coordinate as $n \to \infty$.

This surface state $\bsliv$, called the sliver, has the
property that the left-half and the right-half of the
string are as far as they can be on the unit disk.

\begin{figure}[!ht]
\leavevmode
\begin{center}
\epsfxsize = 15 cm \epsfbox{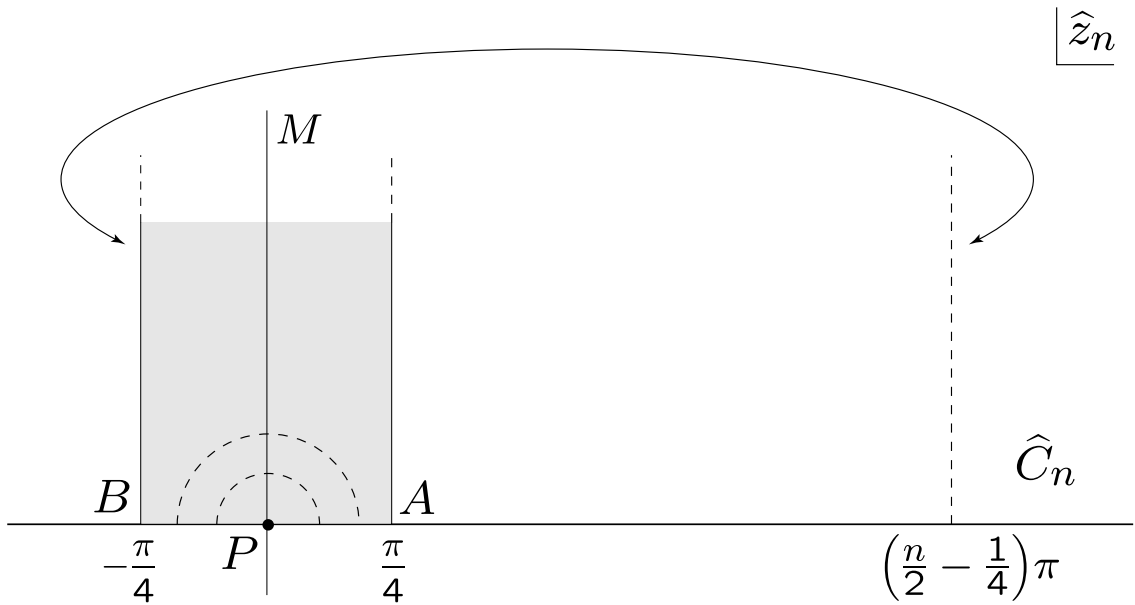 }
\end{center}
\caption[]{\small The finite $n$ approximation
to the sliver presented in the $\wh z_n$ coordinate. 
} \label{f10}
\end{figure}

In the second presentation of the wedge states $\langle n|$, we
represent 
the disk $D$ in a new global coordinate system:
\be \label{ed7}
\gf_n = (w_n)^{n/2}\, .   
\ee
Under this map
the unit disk
$D_0$ in the
$w_n$-coordinates is mapped to a cone in the $\gf_n$ coordinate, 
subtending an
angle $n\pi$ at the origin $\gf_n=0$. We
shall denote this cone by $\wh D_n$.
We see from (\ref{ih}) that
\be \label{ed4}
\gf_n = h(\xi) = {1 + i\xi \over 1 - i\xi}\, .
\ee
Thus $\gf_n$ coordinate system has the special property that
the local coordinate patch, {\it i.e.} the 
image of the half disk $H_U$ in
$\wh D_n$, is particularly simple. 
The image of $H_U$ appears  as a vertical
half-disk of unit radius,
with the curved part of $H_U$ mapped to
the imaginary axis and the diameter 
of $H_U$ mapped to the unit semi-circle
to the right of the imaginary axis.
Using eq.\refb{ed4} we have
\be \label{ed8}
\langle n|\phi\rangle = 
\langle h\circ \phi(0)\rangle_{\wh D_n}\, .
\ee
In the $n\to\infty$ limit $\wh D_n$ can be viewed as an infinite helix.

Finally it is useful to introduce a new coordinate system
\be \label{ecor1}
\wh z_n = {1\over 2 i} \ln \gf_n\, .
\ee
The cone $\wh D_n$ in the $\gf_n$ coordinate system maps to a
semi-infinite cylinder
$\wh C_n$ in the $\wh z_n$ coordinate 
system with $\wh z_n$ spanning the
range:
\be \label{ecor2}
-{\frac{\pi}{4}} \le \Re(\wh z_n) \le (\frac{n}{2}-{1\over 4}) 
\pi\, , \quad
\Im(\wh z_n)\ge 0, \quad \wh z_n \simeq \wh z_n + n\,{\pi\over  2}\, .
\ee
The local coordinate patch 
is the region:
\be \label{ecor3}
-{\pi/4} \le \Re(\wh z_n) \le{\pi/4} , \quad \Im(\wh z_n)\ge 0\, .
\ee
This has been shown in Fig.~\ref{f10}.
The relationship between $\wh z_n$ and 
the local coordinate $\xi$ follows
from eqs.\refb{ed4} and \refb{ecor1}:
\be \label{ecor5}
\wh z_n =\tan^{-1}\xi  \equiv  f(\xi)\, .
\ee
Thus we have
\be \label{ecor6}
\langle n| \phi\rangle = \langle 
f\circ \phi(0)\rangle_{\wh C_n} \quad
\forall
|\phi\rangle\in\HH\, .
\ee
Note that using the periodicity along the $\Re(\wh z_n)$ direction
we
could take the range of $\Re(\wh z_n)$ to 
be $-n\pi/4\le \Re(\wh z_n)\le
n\pi/4$. 
In this case as $n\to\infty$, $\wh C_n$ approaches 
the full UHP and we get
\be \label{esliv}
\langle \Xi |\phi\rangle = \langle 
f\circ \phi(0)\rangle_{UHP} \, .
\ee

\subsubsection{Star multiplication of wedge states} 

The star multiplication of two wedge states 
is easy  to describe by representing them in 
the $\gf_n$ coordinate system. 
In this coordinate system the disk becomes a cone subtending an angle
$\pi n$ at the origin. If we remove the local coordinate patch 
the left over region becomes a sector of
angle $\pi (n-1)$.
The star multiplication of two wedge states
$|m \rangle * |n \rangle$ is readily performed
by gluing
the right half of the sector 
of angle $\pi (m-1)$ to the left half of
the sector of angle $\pi (n-1)$. The result 
is 
of course a sector of angle $\pi (m+n-2)$.
The local coordinate patch must then be restored
to produce the full representation of the surface state.
The result is a cone subtending an 
angle $\pi (m+n-1)$ at the origin, 
which corresponds to the wedge state $|m+n-1 \rangle$
\cite{0006240}:
\be
|m \rangle * |n \rangle = | m+ n -1 \rangle \,.
\ee
In the limit $n,m \to \infty$ we find that the sliver
squares to itself
\be
\Xi  * \Xi =  \Xi \,.
\ee
We can now use the
factorization property to conclude that the matter part
of the sliver obeys $\Xi_m *^m \Xi_m = \Xi_m$,
with 
suitable normalization of $\Xi_m$. (This 
could be
infinite, but is universal in the sense that it 
does not
depend on the specific choice of the matter BCFT, but only
on the value $c=26$ of the central charge).

\medskip
For later use it will be useful to work out the precise relationship
between the different coordinate 
systems appearing in the description of 
the product state
$|m+n-1 \rangle$ and the states $|m \rangle$ and
$|n \rangle$. 
Again this is simple 
in the $\gf$ coordinate
system.
For this let us take
\be \label{ealphamn}
\alpha = \pi(m-1), \qquad \beta=\pi(n-1), \qquad \alpha + \beta =
\pi(m+n-2)\, ,
\ee
and denote by $\RR_\alpha$ 
and $\RR_\beta$ the sectors of angles
$\alpha$ and $\beta$ associated with the states $|m\rangle$ and
$|n\rangle$ respectively.
If we denote by $\gf_m$, $\gf_n$ and $\gf_{m+n-1}$ the
$\gf$ coordinates associated with the wedge states
$|m \rangle$,
$|n \rangle$ and $|m+n-1\rangle$ respectively, we
have
\be \label{ecorpat}  
\gf_{m+n-1} = \cases{ \gf_m \quad \,\,\,\,\,\,
\hbox{in}\quad \RR_\alpha  \cr
 e^{i\alpha} \gf_n \quad \hbox{in}\quad \RR_\beta\,\,. }
\ee
In the $\wh z_n$ coordinate system introduced in eq.\refb{ecor1}
the gluing
relations \refb{ecorpat} take a very  simple form: 
\be \label{ecor4}  
\wh z_{m+n-1} = \cases{ \wh z_m 
\qquad\qquad\qquad \,\,\,\hbox{for} \quad {\pi\over 4}
\le
\Re(\wh z_m) \le (\frac{m}{2}-{1\over 4})\pi, \cr\cr
\wh z_n + {1\over 2} (m-1)\pi \quad \hbox{for} \quad
{\pi\over 4} \le
\Re(\wh z_n) \le (\frac{n}{2}-{1\over 4})\pi\, .}
\ee

\subsection{The sliver as a squeezed state} \label{s3.2}

Here we wish to review the construction of the matter part of the
sliver in the oscillator representation 
and consider the basic algebraic
properties that guarantee that the 
multiplication of two
slivers gives a sliver. In fact we follow the discussion
of Kostelecky and Potting \cite{0008252} who gave the
first algebraic construction of a state that would star
multiply to itself in the matter sector. This discussion
was simplified in \cite{0102112} where due attention was
also paid to normalization factors that guarantee that
the states satisfy precisely the projector 
equation \refb{eo5}.  
We also identify some 
infinite dimensional matrices 
introduced in \cite{0105058} 
with the properties of projection operators.
These matrices will be useful in the construction of multiple D-brane
solutions in section \ref{half}.

\medskip
In order to star multiply two states $|A\rangle$ and $|B\rangle$
we must calculate
\be \label{esstar}
(|A\rangle*|B\rangle)_3  
=  {}_1\langle 
A| {}_2\langle B| V_3\rangle_{123} \,,
\ee
where $|~\rangle_r$ denotes a state in the $r$-th string Hilbert
space, and $|~\rangle_{123}$ denotes a 
state in the product of the Hilbert
space of three strings. 
The key ingredient here is the three-string vertex
$|V_3\rangle_{123}$.  While the vertex has nontrivial
momentum dependence, if the states $A$ and $B$ are at zero 
momentum, the star product gives a zero momentum state that can be
calculated using 
\be \label{e9}
|V_3 \rangle_{123} = 
\exp \Bigl(-  {1\over 2} \sum_{{r,s}}
  a^{(r)\dagger}\cdot V^{rs}
\cdot a^{(s)\dagger}\Bigr)
|0\rangle_{123} \,,
\ee
and the rule $\langle 0 | 0 \rangle =1$. Here the $V^{rs}$,
with $r,s= 1,2,3$, are infinite matrices $V^{rs}_{mn}$ ($m,n=
1,\cdots \infty$) satisfying the cyclicity condition
$V^{rs} = V^{r+1,s+1}$ and the symmetry condition
$(V^{rs})^T = V^{sr}$.  
These properties imply that out of
the nine matrices, three: $V^{11}, V^{12}$ and $V^{21}$, can
be used to obtain all others. 
$a^{(r)\mu\dagger}_m$ ($0\le\mu\le 25$) 
denote oscillators in the $r$-th string Hilbert space. 
For simplicity, the Lorentz and the 
oscillator indices, and the Minkowski matrix
$\eta_{\mu\nu}$
used to contract the Lorentz indices,  
have all been suppressed in eq.\refb{e9}. We shall follow
this convention throughout the paper.

One now introduces  
\be \label{ea1}
M^{rs} \equiv CV^{rs}\,, \quad \hbox{with}\quad
 C_{mn} = (-1)^m \delta_{mn},
\quad m,n\geq 1 \, .
\ee
These matrices can be shown to satisfy the following
properties:
\ben \label{ea222}
&& CV^{rs} = V^{sr} C\,, \quad 
(V^{rs})^T = V^{sr}\, , \nonumber \\ 
&& (M^{rs})^T =
M^{rs}\, , \quad CM^{rs} C 
= M^{sr}\,, \quad [M^{rs}, M^{r's'} ] = 0\, . 
\een
In particular note that all the $M$ matrices commute with each other.
Defining $X \equiv M^{11}$, the three relevant matrices are
$X, M^{12}$ and $M^{21}$. Explicit formulae exist that allow
their explicit 
computation \cite{gross-jevicki,cremmer}.\footnote{Our 
convention \refb{esstar} for describing the star product differs
slightly from that of ref.\cite{gross-jevicki}, the net effect of which
is that the explicit expression for the matrix $V^{rs}$
listed in appendix A of
ref.\cite{0102112}, version 1,  
is actually the expression for the matrix
$V^{sr}$.
Since all the explicit computations performed of ref.\cite{0102112}
involved $V^{sr}$ and $V^{rs}$ symmetrically, this does not affect any
of the calculations in that paper.}
In particular 
they obey the relations:
\be \label{enewrel}
M^{12}+M^{21} = 1 - X\,, \quad 
M^{12} M^{21} = 
X^2 - X\, .  
\ee
 
The state in the matter Hilbert space  
that multiplies to itself  takes
the form \cite{0008252,0102112}
\be \label{e21a}
|\Psi\rangle = \NN^{26}    
\exp \Bigl(-{1\over 2}\, a^\dagger \cdot S\cdot a^\dagger
\Bigr) |0\rangle\,, \quad \NN= 
\{\det(1 - X)\det (1+T) \}^{1/2}\,, \quad
S = CT
\ee
where the matrix $T$  satisfies $CTC= T$ and the
equation
\be \label{ea44}
XT^2 - (1+X) T + X=0\, ,
\ee
which gives
\be \label{ea44a}
T = (2 X)^{-1} ( 1 + X -
\sqrt{(1+3 X) (1 - X)}). 
\ee
In taking the square root we pick that branch which,
for small $X$, goes as $(1+X)$.

In \cite{0102112} we identified this state as the sliver by
computing numerically the matrix $S$ using the equation
above and comparing 
the state obtained this way with the matter part of
the sliver $|\Xi\rangle$, which can be evaluated directly 
using the techniques of ref.\cite{LPP}.   
We found close agreement between the numerical
values
of $S_{mn}$ and exact answers for $\wh S_{mn}$.
This gave convincing evidence that $|\Psi\rangle = |\Xi\rangle$.

\medskip
The algebraic structure allows one to construct
a pair of projectors that have the interpretation
of projectors into the left and right halves of
a string. We define the matrices 
\ben
\label{proj}
\rho_1 &&= {1\over (1+T)
(1-X)}\Bigl[ M^{12} (1-TX)  + T (M^{21})^2  \Bigr] \,, \\
\rho_2 && =  {1\over (1+T)
(1-X)} \Bigl[ M^{21} (1-TX) + T (M^{12})^2\Bigr] \,. \nonumber
\een 
One can verify,
using various identities  
satisfied by $X$, $M^{12}$ and $M^{21}$,
that they satisfy the following properties:
\be
\rho_1 + \rho_2 = 1, \quad \rho_1^T 
= \rho_1\,, \quad \rho_2^T = \rho_2\,,
\quad C\rho_1 C =
\rho_2\,,
\ee
and more importantly
\be  \label{erhoproj}  
\rho_1 \rho_1 = \rho_1\,, \quad \rho_2 \rho_2 = \rho_2\,.
\ee
We see that $\rho_1$ and $\rho_2$ are
projection operators into orthogonal complementary subspaces 
exchanged by $C$.

With the help of these projectors one can compute
some useful star products.  For example, introducing 
coherent like
states of the form
\be \label{eb2}
|\Xi_\beta \rangle =
\exp \Bigl( \sum_{n=1}^\infty
(-)^{n+1}\beta_{\mu n} a_n^{\mu \dagger}\Bigr) |\Xi
\rangle =
\exp ( -a^\dagger \cdot C\beta ) |\Xi
\rangle\, ,  
\ee
one can show \cite{0105058} that 
\be \label{eb9}
|\Xi_{\beta_1}\rangle * |\Xi_{\beta_2}\rangle 
= \exp \Bigl(  -
{\cal C} (\beta_1, \beta_2) \Bigr)  |\Xi_{\rho_1\beta_1
+\rho_2\beta_2}\rangle \, ,
\ee
where
\ben \label{eb7}
{\cal C} (\beta_1, \beta_2) 
={1\over 2}\, (\beta_1, \beta_2){1\over (1+T)(1-X)} 
\pmatrix{V^{11}(1-T) & V^{12} \cr V^{21} & V^{11}(1-T) }
\pmatrix{\beta_1 \cr \beta_2}  \,. 
\een
\refb{eb9}  is a useful relation that allows one to 
compute $*$-products  
of slivers acted by oscillators by simple differentiation.
For example, 
\ben \label{eb11} 
 (a^{\mu\dagger}_{m}|\Xi\rangle) *  
(a^{\nu\dagger}_{n} |\Xi\rangle) 
=(-1)^{(m + 1) + (n + 1)}  \Bigl( {\partial \over \partial 
\beta_{1\mu m}}  
 {\partial \over \partial \beta_{2\nu n}} 
\big(|\Xi_{\beta_1}\rangle *
|\Xi_{\beta_2}\rangle\big)\Bigr)_{\beta_1=\beta_2=0}\, .
\een

\medskip
\noindent
It follows from \refb{eb9} that 
$P_\beta \equiv \exp \Bigl(  
{\cal C} (\beta, \beta) \Bigr)  |\Xi_\beta\rangle$
satisfies $P_\beta * P_\beta = P_\beta\,$.
One can also check that
$\langle P_\beta | P_\beta\rangle = \langle\Xi | \Xi\rangle\, $.

\sectiono{Boundary CFT construction of 
D-brane solutions} \label{sbound}

In this section we  shall review the basic 
construction
of \cite{0105168}. We shall describe
deformations of the sliver to generate 
new solutions of the
equations of motion 
representing D-branes described by general BCFT's.
The general idea is simple.
We denote by BCFT$_0$ the {\it reference} BCFT  
in whose Hilbert space the
string field takes value.
The D-brane associated with BCFT$_0$ is represented by a
solution of the string field equation \refb{eo2} whose
matter part is
the sliver of BCFT$_0$, 
$-$ the surface  state described in section \ref{sursl},
with the  specific boundary 
condition corresponding to BCFT$_0$ on
the boundary of the surface.
To get a solution representing the single D-brane 
of  some other boundary conformal field theory 
BCFT$'$ we must represent the sliver of BCFT$'$ on the
state space of BCFT$_0$, as we now explain. 
The construction assumes that BCFT$_0$ and BCFT$'$ have the same bulk
conformal field theory,  but of course, differ  in
their boundary interactions.

 Usually the sliver $\Xi'$ of BCFT$'$ will be
described in the same manner as 
discussed in the last section, with all
the correlation functions now 
being calculated in BCFT$'$. This, however,
would express $\Xi'$ as a state in the state 
space $\HH'$ of BCFT$'$, 
since in eq.~\refb{esliv}, 
for example, $\phi$ will now represent a
vertex  
operator of BCFT$'$. In order to express $\Xi'$ in the 
state  space of
BCFT$_0$, we adopt the following procedure.
As discussed in section \ref{sursl1}, at 
an intuitive
level, the wave-functional for $\langle\Xi'|$ is a 
functional of $\vp_0(\sigma)$
with $\sigma$ 
labeling the coordinate along the string,
obtained by 
functional
integration over the two dimensional 
fields $\vp$ on the full disk (UHP
labeled by $z'=\tan^{-1}\xi$  
in this
case) minus the local coordinate patch
($|\Re(z')|\le \pi/4$), with boundary condition
$\vp=\vp_0(\sigma)$ on the 
boundary $\xi=e^{i\sigma}$ 
($\Re(z')=\pm\pi/4$), and the boundary condition
appropriate to BCFT$'$ on the rest of the 
boundary of this region ($z'$
real,
$|\Re(z')|\ge \pi/4$). 
On the other hand given a state $|\phi\rangle$ in  
the Hilbert space $\HH$ of BCFT$_0$, 
we represent
the
wave-functional of
$|\phi\rangle$ as a functional of $\vp_0(\sigma)$,  obtained by
performing path
integral over $\vp$ in the inside of the  local coordinate patch {\it
with
boundary
condition appropriate for BCFT$_0$} on the real axis, 
the 
vertex operator 
$\phi$ of BCFT$_0$ 
inserted at the origin in the local coordinate system $\xi$, and 
the boundary condition $\vp=\vp_0(\sigma)$ on the 
semi-circle $\xi=e^{i\sigma}$.  
Finally in order to calculate $\langle\Xi'|\phi\rangle$ we
take
the product of the wave-functional of $\langle\Xi'|$
and the wave-functional of $|\phi\rangle$ and integrate over
$\vp_0(\sigma)$. The result will be a functional integral
over the fields $\vp$ on the full UHP, with boundary condition
appropriate to BCFT$'$ in the range $|\Re(z')|\ge \pi/4$,
boundary condition corresponding to BCFT$_0$    
for $|\Re(z')|\le \pi/4$ and the 
vertex operator
$\phi$ inserted at the origin {\it in the local coordinate system}.
This can be expressed as
\be \label{eexpress}
\langle\Xi'|\phi\rangle = \langle f\circ\phi(0)\rangle''\, ,
\ee
where $\langle ~\rangle''$ denotes  
correlation function in a theory where
we have BCFT$'$ boundary condition 
for $|\Re(z')|\ge \pi/4$ and BCFT$_0$
boundary condition for $|\Re(z')|\le \pi/4$. 
$\phi$ is a vertex operator
in BCFT$_0$ and $f(\xi)=\tan^{-1}\xi$ as usual.

In what follows, we shall show that after appropriate ultraviolet 
regularization,  
$\Xi'$ defined this way squares to   
itself under $*$-multiplication, and also has the right tension for
describing a D-brane associated to BCFT$'$. 
We will only review this general case in 
the next subsections. We refer to \cite{0105168}
for the discussion of the case where
BCFT$'$ is replaced by a general two
dimensional field theory obtained from BCFT$_0$ by some boundary 
perturbation.

\subsection{Solution describing an arbitrary D-brane} \label{sn1}

We shall now describe the construction of the solution of the SFT 
equations of motion describing a D-brane 
corresponding to an 
arbitrary boundary conformal field theory BCFT$'$ with 
the same bulk  
CFT as BCFT$_0$.
We start with the definition of $\Xi'$ given in eq.\refb{eexpress}.
The effect of the change of the 
boundary condition beyond $|x'|\ge \pi/4$
can be represented by inserting boundary condition
changing vertex operator $\sigma^\pm$ (discussed {\it e.g.} in
\cite{cardy, 9704006}) 
at $x'=\pm\pi/4$. In other words we can express 
\refb{eexpress} as
\be \label{exy1}
\langle\Xi'|\phi\rangle = \Bigl\langle f\circ\phi(0) 
\, \sigma^+({\pi\over 4}) \sigma^-(-{\pi\over 4}) \Bigr\rangle\, .
\ee
If we denote by D and D$'$ the  D-branes
associated with BCFT$_0$ and BCFT$'$ respectively, then $\sigma^+$
denotes 
the vertex operator for the ground state of an open string whose left
end  (viewed from inside the UHP) 
is  on the D$'$-brane 
and right end is on the D-brane, whereas $\sigma^-$ denotes the vertex 
operator  for the ground state of open string whose left end is     
on the D-brane and right end is on the D$'$-brane. 
In anticipation of  short-distance divergences, we shall actually put
$\sigma^-$ and $\sigma^+$ at 
$(-{\pi\over 4}-\epsilon)$ and $({\pi\over 4}+\epsilon)$ respectively, 
where $\epsilon$ is a small positive number. We 
shall also use the description of the sliver as limit of finite $n$
wedge 
states in the $\wh z_n$ coordinate 
introduced in eq.\refb{ecor1}. Thus we have
\be \label{exy2}
\langle\Xi'|\phi\rangle =   \lim_{n\to\infty} 
\Bigl\langle f\circ\phi(0)
\, \sigma^+\Bigl({\pi\over 
4}+\epsilon\Bigr) 
\sigma^-\Bigl({n\over 2}\pi-{\pi\over 4}-\epsilon\Bigr)
\Bigr\rangle_{\wh  C_n}\, .
\ee   
We now 
calculate $\Xi'*\Xi'$.
{}From the gluing relations \refb{ecor4} we get,
\ben \label{exy3}  
\langle \Xi' * \Xi'|\phi\rangle &=& \lim_{m,n\to\infty}  \Biggl\langle 
f\circ
\phi(0) \,\, \sigma^+\Bigl( {\pi\over 4} +\epsilon\Bigr)
\,\sigma^-\Bigl( ({m\over 2}-{1\over 4})\pi 
-\epsilon\Bigr) 
\nonumber \\
&&  \sigma^+\Bigl(
({m\over
2}-{1\over 4})\pi+\epsilon\Bigr) 
\sigma^-\Bigl(
{1\over 2} (m+n-1) \pi -{\pi\over 4} 
-\epsilon\Bigr)\Biggr\rangle_{\wh C_{m+n-1}} \, . 
\een
Thus the $\sigma^\pm$ at ${1\over 2}
(m-{1\over  2})\pi 
\pm\epsilon$
come close as $\epsilon\to 0$ and give rise to a divergent factor 
$(2\epsilon)^{-2h}$ where $h$ is the conformal weight of $\sigma^\pm$.
Hence we have $\Xi' * 
\Xi'= (2\epsilon)^{-2h}\Xi'$. 
This requires us to
redefine $\Xi'$ by absorbing a factor of $(2\epsilon)^{2h}$,
so that it squares to itself under $*$-multiplication
\be 
\label{rt}
\Xi'_\rr 
\equiv (2\epsilon)^{2h}\,\, \Xi'  \quad \to \quad
\Xi'_\rr * \Xi'_\rr = \Xi'_\rr \,.
\ee
We note, however, that even for finite $\epsilon$, the state
$\Xi'_\rr$ still squares to itself. 
Indeed, the
product 
$\sigma^-( {1\over
2}(m-{1\over 2})\pi -\epsilon) \sigma^+({1\over  2}(m-{1\over
2})\pi + \epsilon) $ 
can be expanded in an operator product 
expansion,
and since these operators are moved to $\infty$ in the $m,n\to\infty$
limit, only the identity operator in this operator product expansion 
would contribute. 
The coefficient of the identity operator is given by 
$(2\epsilon)^{-2h}$ 
even for finite $\epsilon$.

Since BCFT$_0$ and BCFT$'$ differ 
only in the matter sector, it is clear 
that 
$|\Xi'_\rr\rangle$ has the factorized form
\be \label{epfac}
|\Xi'_\rr\rangle=|\Xi_g\rangle \otimes |\Xi'_{\rr,m}
\rangle\, ,
\ee
where $\Xi_g$ is a universal ghost factor. Normalizing
$\Xi_g$ 
(which is independent of 
the choice of BCFT$'$) 
such that $\Xi_g*^g\Xi_g=\Xi_g$,
we can
ensure that
\be \label{eprodm} 
\Xi'_{\rr,m} *^m \Xi'_{\rr, m} = \Xi'_{\rr,m}\, .
\ee
Thus we can now construct a new D-brane solution by taking the product
$|\Psi_g\rangle\otimes |\Xi'_{\rr,m}\rangle$, where 
$|\Psi_g\rangle$
is the same universal ghost state that appears 
in the construction of the
D-brane solution corresponding to BCFT$_0$.

We shall now calculate the tension associated with this new D-brane 
solution. For this we need to compute
$\langle
\Xi_{\rr,m}'|\Xi_{\rr,m}'\rangle_m$, where the 
subscript $m$ denotes matter. 
We have,
\be \label{exy4}
\langle\Xi_{\rr,m}'|\Xi_{\rr,m}'\rangle_m = (2\epsilon)^{4h}\,\,
\langle\Xi_{m}'|\Xi_{m}'\rangle_m \,.
\ee
Calculation of $\langle\Xi_{m}'|\Xi_{m}'\rangle_m$  is again simple
in the $\wh z_n$
coordinate system. We first compute the $*$-product of the two states,
and then in the final glued surface with coordinate $\wh z_{m+n-1}$ we
remove the 
local coordinate patch 
$-\pi/4\le \Re(\wh
z_{m+n-1})\le \pi/4$ and identify the lines $\Re(\wh z_{m+n-1})=\pm
\pi/4$. This produces the semi-infinite cylinder $\wt C_{m+n-2}$
defined by  
${\pi\over 4} \leq \Re(
z_{m+n-2}) \leq {\pi\over 4}+ (m+n-2){\pi\over 2}$ and $\Im (z_{m+n-2})
\geq 0$. 
We therefore find 
\ben \label{exy5}
\langle\Xi_{m}'|\Xi_{m}'\rangle_m 
&=& \lim_{m,n\to\infty}  \Biggl\langle 
\sigma^+\Bigl( {\pi\over 4} +\epsilon\Bigr)\,\,
\sigma^-\Bigl( ({m\over 2}-{1\over 4})\pi 
-\epsilon\Bigr) 
\nonumber \\
&&  \sigma^+ \Bigl( {m\over
2}-{1\over 4})\pi + \epsilon\Bigr)  
\sigma^-\Bigl(
{1\over 2} (m+n-1) \pi -{\pi\over 4} 
-\epsilon\Bigr)\Bigg\rangle_{\wt C_{m+n-2}} \, , 
\een
where the correlation function is now being computed in the 
matter BCFT.
We get 
a factor of
$(2\epsilon)^{-2h}$ 
coming from $\sigma^\pm$ inserted at $({m\over 2}-{1\over 4})\pi  
\pm\epsilon$ as before, but there is 
another factor of $(2\epsilon)^{-2h}$
coming from the other $\sigma^\pm$ insertions that happen at points
separated by a (minimal) distance $2\epsilon$ on the cylinder
$\wt C_{m+n-2}$.
These exactly cancel the explicit
factor of
$(2\epsilon)^{4h}$ in \refb{exy4}.
Since from the definition 
of $\sigma^\pm$ it is clear that in the $\epsilon\to 
0$ limit we have BCFT$'$ boundary condition on the full real 
$z_{m+n-2}$ 
axis, we 
find that $\langle \Xi'_{\rr,m} |
\Xi_{\rr, m}'\rangle_m$ is given by the partition function of the
deformed boundary CFT on the $\wt C_{m+n-2}$  cylinder 
\be \label{exy6}
\langle
\Xi_{\rr,m}'|\Xi_{\rr,m}'\rangle_m = Z_{\wt C_{m+n-2}}
 (\hbox{BCFT}')
\sim Z_{D_0} (\hbox{BCFT}')\,,
\ee
where in the last step we relate this partition function to
the one on the standard unit disk. This is possible 
because of conformal
invariance. 
Any constant multiplicative factor that might appear due to conformal 
anomaly depends only on the bulk central charge and is independent of
the 
choice of BCFT$'$. This can at most give rise to a universal
multiplicative 
factor.
Since the partition function of BCFT$'$ 
on the unit disk is proportional
to  the tension of the corresponding 
D-brane~\cite{TEN} 
$-$ a fact  
which has played a crucial role in the analysis of tachyon
condensation in boundary string field theory\cite{BOLD,BNEW}, $-$
we see that the tension $\langle
\Xi_{\rr,m}'|\Xi_{\rr,m}'\rangle_m$ computed from  vacuum 
string field theory agrees with the known tension of the BCFT$'$
D-brane, up to an overall 
constant factor independent of BCFT$'$. 

\medskip
Arguments similar to the 
one given for $\Xi'*\Xi'$ show 
that the 
result \refb{exy6} holds even when $\epsilon$ is finite. 
In this case we 
have two pairs 
of $\sigma^\pm$ on the boundary, with the first pair being 
infinite 
distance away from the second pair. Thus we can expand each pair using 
operator product expansion 
and only the identity operator contributes, 
giving us back the partition function of BCFT$'$ on the disk.
{}From this we see that we have a one parameter family of solutions, 
labeled by $\epsilon$, describing the same D-brane. We expect these 
solutions to be related by gauge 
transformations\footnote{$\p_\epsilon\Xi'_{\rr,m}$ 
has finite norm (as can be easily
verified) and hence is pure gauge according to the arguments  
given in section 5 of \cite{0105168}.}.

\subsection{Multiple D-branes and coincident D-branes}
\label{snew4.2}

We first consider the construction of a configuration
containing various D-branes associated to different
BCFT$'$s.
To this end, we note that the star product $\Xi'*\Xi$ of
the BCFT$'$ solution and the BCFT$_0$ solution vanishes.
Indeed, using the same methods as in the previous
subsection, the computation of 
$\langle\Xi'*\Xi|\phi\rangle$  
leads to the cylinder $\wh C_{m+n-1}$ with a
$\sigma^+$ insertion at
${\pi\over 4}+\epsilon$,
a $\sigma^-$ insertion at $({m\over 2} -{1\over
4})-\epsilon$,
and $f\circ\phi(0)$ insertion at $f(0)=0$. 
In
the  $m,n\to\infty$ limit, $\sigma^-$ moves 
off to infinity and as a result
the  correlation function vanishes since $\sigma^-$ has 
dimension larger than zero {\em   as 
long as BCFT$_0$ and BCFT$'$ are different.} Similar arguments 
show that 
$\Xi*\Xi'$ and $\langle\Xi_m|\Xi'_m\rangle_m$ also vanish. Thus 
the matter part  of $\Xi+\Xi'_\rr$ is a new solution  
describing the superposition of the D-branes corresponding  to 
BCFT$_0$ and BCFT$'$. 
Since no special assumptions were made about BCFT$_0$ 
or BCFT$'$, it follows that $\Xi'*\Xi''=\Xi''*\Xi' = 0$ and 
$\langle\Xi'_m|\Xi''_m\rangle_m =0$ for any two different 
BCFT$'$ and BCFT$''$, and hence we can superpose any number 
of slivers to form a solution. This in 
particular also includes theories which differ from each other by a
small marginal deformation. Special cases of 
this phenomenon, in the case of D-branes in flat space-time, have been 
discussed in ref.\cite{0105059}.

\medskip
This procedure, however, is not suitable for superposing
D-branes 
associated with the same BCFT, {\it i.e.} for parallel coincident 
D-branes. For example,
 if we take BCFT$'$ to differ from BCFT$_0$ by
an  exactly 
marginal deformation with deformation parameter $\lambda$, 
then in the 
$\lambda\to 0$ limit the operators $\sigma^\pm$ 
both 
approach the identity operator (having vanishing conformal weight),
and although the argument
 for the  vanishing of $\Xi*\Xi'$ holds for any non-zero $\lambda$,
it breaks down at $\lambda=0$.

In order to construct a superposition of identical D-branes, one can
proceed in a different way. 
First consider getting coincident BCFT$_0$ branes. To this
end we introduce a modified BCFT$_0$ sliver
\be \label{exy222}
\langle\Xi_\chi|\phi\rangle =   \lim_{n\to\infty} 
\Bigl\langle f\circ\phi(0)
\, \chi^+\Bigl({\pi\over 
4}+\epsilon\Bigr) \chi^-\Bigl({n\over 2}\pi-{\pi\over 4}-\epsilon\Bigr)
\Bigr\rangle_{\wh  C_n}\, .
\ee   
Here  $\chi^\pm$ are a conjugate pair\footnote{We
need to choose $\chi^\pm$ to be conjugates of each other so that
the string field is hermitian.} of operators of BCFT$_0$,
having a common dimension $h$ greater than zero, and representing some 
excited states of the open string with
{\it both
ends having BCFT$_0$ boundary condition}. Thus, throughout the
real line we have BCFT$_0$ boundary conditions. We require
that the coefficient of the identity in the 
OPE $\chi^-(x) \chi^+(y)$ is
given by $ |x-y|^{-2h_i}$, and that this 
OPE does not
contain any  other operator of dimension $\le 0$.   

The clear parallel between eqn.~\refb{exy222} and eqn.~\refb{exy2},
describing the BCFT$'$ D-brane, implies that an  analysis
identical to
the one carried out in the previous section will show that:
\begin{enumerate}
\item This new state $\Xi_\chi$ (after suitable renormalization as in
eq.\refb{rt}) squares to itself under
$*$-multiplication.
\item The BPZ norm of the matter part of 
$\Xi_\chi$ is proportional to the
partition
function of BCFT$_0$ on the unit disk.
\item $\Xi_\chi$ has vanishing $*$-product with $\Xi$.
\end{enumerate}
Thus the matter part of this state gives another representation of the
D-brane associated with BCFT$_0$, 
and we can construct a pair of D-branes
associated with BCFT$_0$ by superposing the matter parts of $\Xi$ and
$\Xi_\chi$. 

This construction
can be easily generalized to describe 
multiple BCFT$_0$ D-branes. 
We construct different
representations of the
same D-brane by
using different vertex operators $\chi^{(i)\pm}$ in
BCFT$_0$ satisfying the `orthonormality condition' that
the coefficient of the identity operator in the OPE of
$\chi^{(i)-}(x) \chi^{(j)+}(y)$ is given by
$\delta_{ij} |x-y|^{-2h_i}$, and  that this 
OPE does not
contain any other operator of dimension $\le 0$. 
The correponding
solutions
$\Xi_{\chi^{(i)}}$ all have vanishing $*$-product with each other,
and hence can be superposed to represent multiple D-branes
associated
with BCFT$_0$.

For constructing a general configuration of multiple D-branes some
of which
may be identical and some are different, we choose a set of
conjugate pair of
vertex operators $\chi^{(i)\pm}$, representing open strings
with one end satisfying boundary condition corresponding to
BCFT$_0$ and the other end satisfying the boundary condition
corresponding to some boundary conformal field theory BCFT$_i$,
satisfying the `orthonormality
condition' that
the coefficient of the identity operator in the OPE of
$\chi^{(i)-}(x) \chi^{(j)+}(y)$ is given by
$\delta_{ij} |x-y|^{-2h_i}$, and  that this 
OPE does not
contain any other operator of dimension $\le 0$. 
The correponding
solutions
$\Xi_{\chi^{(i)}}$ all have vanishing $*$-product with each other,
and hence can be superposed to represent multiple D-branes, with
the $i$th D-brane being
associated
with BCFT$_i$. Since there is no restriction that BCFT$_i$
should be different  from BCFT$_0$, or from BCFT$_j$ for $j\ne i$,
we can use this procedure to describe superposition of an arbitrary
set of D-branes.

This procedure of adding vertex 
operators near $\pm\pi/4$ to create new
solutions representing the same D-brane is
the BCFT version 
of the use of excited states of half-strings \cite{0105058,0105059} 
for the same purpose. 
This is reviewed in section \ref{half}.

\subsection{Solutions from boundary field theories}
\label{newsection}  

We can consider a class of solutions associated with the 
sliver for boundary field theories which are not necessarily 
conformal. For this,  suppose $V$ is a local vertex 
operator in the matter 
sector  of BCFT$_0$ and define a new state $\langle
\Xi^{V,\lambda}|$  through the
relation:\footnote{A construction 
that is similar in spirit but uses a
different geometry was suggested in ref.\cite{0011009}.}
\be \label{epp2}
\langle \Xi^{V,\lambda}| \phi\rangle = \lim_{n\to\infty} \Big\langle 
\exp\Big(-\lambda \int_{{\pi\over 4}}^{({n\over 2} -{1\over
4})\pi} V(x_n) dx_n\Big) f\circ
\phi(0)\Big\rangle_{\wh C_n} \quad
\forall |\phi\rangle\in\HH\, ,
\ee
where $x_n=\Re(\wh z_n)$, $\lambda$ is a constant, 
and the integration is
done over the real $\wh z_n$ axis {\it excluding the part that is inside
the local coordinate patch.}  
This expression should be treated  
as a correlation function in a theory where on part of the boundary we
have the usual boundary action corresponding to 
BCFT$_0$, 
and on part
of the boundary we have a modified boundary action obtained by adding
the integral of $V$ to the original action (in defining this we need
to use
suitable  regularization and renormalization  
prescriptions; see ref.\cite{0105168} for more discussion of
this).
Alternatively, we have a
correlation function with BCFT$_0$ 
boundary condition in the range
$-{\pi\over 4}\le x\le {\pi\over 4}$ and a modified boundary condition
outside this range.

One can show that $|\Xi^{V,\lambda}\rangle$ satisfies the projection 
equation $\Xi^{V,\lambda} * \Xi^{V,\lambda} = \Xi^{V,\lambda}$,
by using the $\wh z_n$ coordinate system to take the star product
\cite{0105168}. This may be
surprising  given that $V$ is not constrained, but  is
a consequence of the trivial way  the star product acts in the $\wh
z_n$ coordinates .
Since the operator $V$ is in the matter sector, 
$\Xi^{V,\lambda}$ has the usual ghost/matter
factorized form, and the matter part satisfies
$\Xi^{V,\lambda}_m *^m \Xi^{V,\lambda}_m = \Xi^{V,\lambda}_m\,$.
Thus we can now construct new D-brane solutions by taking the product
$|\Psi_g\rangle\otimes |\Xi^{V,\lambda}_m\rangle$, where
$|\Psi_g\rangle$ 
is the 
universal ghost state that appears in the 
D25-brane solution.

\medskip
The tension associated with such solution
is proportional to $\langle
\Xi_m^{V,\lambda}|\Xi_m^{V,\lambda}\rangle_m$.
This computation is again simple
in the $\wh z_n$ coordinates 
and the relevant geometry was discussed
above \refb{exy5}. One finds
\be \label{epp6}
\langle \Xi_m^{V,\lambda}|\Xi_m^{V,\lambda}\rangle_m =
\lim_{m,n\to\infty}  \Bigl\langle 
\exp\Big(-\lambda \int_{ {\pi\over 4}}^{{m+n-1\over
2}\pi -{\pi\over 4}} V(x_{m+n-2})
dx_{m+n-2}\Big)
\Bigr\rangle_{\wt C_{m+n-2}} \, , \nonumber \\
\ee
where $x_{m+n-2}=\Re(z_{m+n-2})$. 
We now define a rescaled coordinate $u$ as 
$u = 4(z_{m+n-2}-{\pi\over 4})/(m+n-2)$   
so that $\Re(u)$ 
ranges from 0 to $2\pi$. Thus in the
$u$ coordinate we have a semiinfinite cylinder $C$ of circumference
$2\pi$.
Writing $u=i\rho+\theta$, and 
taking into account the conformal transformation of the vertex
operator $V$ under this scale transformation, 
we get:
\be \label{epp7}
\langle \Xi_m^{V,\lambda}|\Xi_m^{V,\lambda}\rangle =
\lim_{m,n\to\infty}  \Bigl\langle
\exp\Big(-\lambda_R \int_{\theta=0}^{2\pi}
d \theta V_R(\theta) \Big) \Bigr\rangle_{C}\, ,
\ee
where $\lambda_R\int V_R$ denotes the operator to which the 
perturbation $\lambda \int V$ flows under the rescaling by 
$(m+n-2)/4$. 
This semiinfinite cylinder in the $u$ coordinate is
nothing but a unit disk $D_U$ with $\theta$ labeling the angular
parameter along the boundary of the disk, and $e^{-\rho}$ 
labeling the radial coordinate. Thus \refb{epp7} is 
the partition
function on a unit disk, with the perturbation 
$\lambda_R \int V_R(\theta) d\theta$
added at the boundary!
If $V$ is 
 a relevant deformation then $h<1$
and 
in the limit $m,n\to\infty$, $\lambda_R\int d\theta V_R(\theta)$ 
approaches its infrared fixed point $\lambda_{IR} \int d\theta 
V_{IR}(\theta)$.\footnote{Irrelevant perturbations flow to zero in the
IR, 
and are not expected to  give rise to new solutions.}   Thus,
$\langle \Xi_m^{V,\lambda}|\Xi_m^{V,\lambda}\rangle_m$ represents the
partition
function on the unit disk of the BCFT to which 
the theory flows in the
infrared! This 
is indeed the tension of the D-brane associated to this BCFT. 
Thus  $|\Psi_g\rangle\otimes |\Xi_m^{V,\lambda}\rangle$ 
is the D-brane solution
for the 
BCFT obtained as the  infrared fixed point 
of the boundary perturbation  $\int V(\theta) d\theta$. 
\footnote{The solution \refb{epp2} does seem to depend
on $\lambda$ for a general relevant perturbation. Since different
values of $\lambda$ correspond to the same 
tension of the final brane, we
expect that they represent gauge equivalent solutions. The
parameter $\lambda$
is analogous to the parameter $b$ labeling the lower
dimensional D-$p$-brane solutions considered in
ref.~\cite{0102112} (this was suggested to us by
E.~Witten.).}

\sectiono{Half strings, projectors and multiple D25-branes in the
algebraic approach}
\label{half}

In this section we review the basic
ideas of \cite{0105058} $-$ see also \cite{0105059}. 
First we consider the functional representation
of the $*$-product and argue that it is natural to think
of string fields as operators acting in the space of half string
functionals. This intuition provides the clue
for a rigorous algebraic construction of multiple
D-brane solutions.  
Throughout this section we shall deal with matter sector states
only, and compute $*$-products and BPZ inner products in the
matter sector, but will drop the subscripts and superscripts $m$
from the labels of the states, $*$-products and inner products.

\subsection{Half-string functionals and projectors}  

\newcommand{\sm}{*^m}

We shall begin by examining the representation of
string fields as functionals of half strings.
  This viewpoint
is possible at least for the case of zero momentum string fields.
It leads to the realization that the sliver functional 
factors into functionals of the left and right halves of the string,
allowing its interpretation as a rank-one projector in the
space of half-string functionals. We construct higher rank
projectors -- these are 
solution of the equations of motion representing multiple
D25-branes.

\subsubsection{Zero momentum string field as a matrix} \label{smds1}

The string field equation in the matter sector is given by 
\be \label{eg1}  
\Psi * \Psi = \Psi\, .
\ee
Thus if we can regard the string field as an operator acting on
some
vector space where $*$ has the interpretation of product of operators,
then $\Psi$ 
is a
projection operator in this vector space. 
Furthermore, in analogy with
the results in
non-commutative solitons \cite{noncom} 
we expect that in
order to describe a single
D-brane, $\Psi$ 
should be a projection operator into a single
state in this vector space.

A possible operator interpretation of the 
string field was suggested in
Witten's original paper \cite{WITTENBSFT}, and was further
developed in
refs.\cite{comma,BORDES}.  
In this picture the string field is viewed  as a
matrix where the role of the row index and the
column index are taken by the left-half and the 
right-half of the string respectively.
In order
to make this more concrete, let us consider 
the standard mode expansion of
the open string coordinate \cite{gross-jevicki}: 
\be \label{eg2}
X^\mu(\sigma) = x^\mu_0 + \sqrt 2 \sum_{n=1}^\infty x^\mu_n
\cos(n\sigma)\,,\quad \hbox{for}\quad 0\le\sigma\le
\pi\, . 
\ee
Now let us introduce coordinates
$X^{L\mu}$
and $X^{R\mu}$ for the left and the right half of the string as 
follows\footnote{Here the half strings are parameterized
both from $\sigma=0$ to $\sigma=\pi$, as opposed to
the parameterization of \cite{BORDES} where the half strings
are parameterized from $0$ to $\pi/2$.
}:
\ben \label{eg3}
X^{L\mu}(\sigma) &=& X^\mu(\sigma/2) - 
\,\,X^\mu(\pi/2)\,\,, \,\,\qquad \hbox{for}\quad 0\le\sigma\le
\pi\, . \nonumber\\
X^{R\mu}(\sigma) &=&
X^\mu(\pi - \sigma/2) - X^\mu(\pi/2), 
\qquad \hskip-9pt\hbox{for}\quad 0\le\sigma\le
\pi\, .
\een
$X^{L \mu}(\sigma)$ and $X^{R \mu}(\sigma)$ satisfy the usual Neumann
boundary condition at $\sigma=0$ and a Dirichlet boundary condition at
$\sigma=\pi$. Thus they have expansions of the form:
\ben \label{eg4}
X^{L\mu}(\sigma) &=& \sqrt 2 \sum_{n=1}^\infty x^{L\mu}_n
\cos((n-{1\over 2})\sigma),  \\
 X^{R\mu}(\sigma) &=& \sqrt 2
\sum_{n=1}^\infty x^{R\mu}_n
\cos((n-{1\over   
2})\sigma)\, .\nonumber
\een
Comparing \refb{eg2} and \refb{eg4} we get an expression for
the full open string modes in terms of the modes of the left-half
and the modes of the right-half:
\be \label{eg5}
x^\mu_n = A_{nm}^+ \,x^{L\mu}_m +  A^-_{nm} \, x^{R\mu}_m, \qquad
m,n\ge 1\, ,
\ee
where the matrices $A^\pm$ are  
\be \label{eg7}
A_{nm}^\pm =\pm{1\over 2} \delta_{n, 2m-1} +{1\over 2\pi} 
\epsilon (n,m) \Big({1\over 2 m + n - 1} +
{1\over 2 m - n - 1}\Big) \, ,
\ee
and 
\be   \epsilon (n,m) =
(1 + (-1)^n) (-1)^{m+{1\over 2} n - 1} \,.
\ee
Alternatively we can write the left-half modes and right
half modes in terms of the full string modes
\ben \label{eg6}
x^{L\mu}_m &=& \wt A_{mn}^+ \,x^\mu_n, \\
 x^{R\mu}_m &=& \wt
A_{mn}^-\, x^\mu_n,
\qquad m,n\ge 1\, , \nonumber
\een
where one finds
\be \label{eg8}
\wt A_{mn}^\pm  =2 A_{nm}^\pm  -{1\over \pi}\, \epsilon(n,m)\,
\Big( {2\over 2 m -1}\Big) \, . 
\ee
Note that
\be \label{egtwist}
A^+ = C A^-, \quad \wt A^+ = \wt A^- C\, ,
\ee
where $C_{mn}=(-1)^{n} \delta_{mn}$ is the twist operator.
Note also that the relationship between $x_n^\mu$ and $(x_n^{L\mu}$,
$x_n^{R\mu})$ does not involve the zero mode $x_0^\mu$ of $X^\mu$.

A general string field configuration can be 
regarded as a functional of
$X^\mu(\sigma)$, or equivalently a function of the infinite set of
coordinates $x^\mu_n$. Now suppose we have a translationally invariant
string field configuration. In this case 
it is independent of $x_0^\mu$
and we can regard this as a function 
$\psi(\{x^{L\mu}_n\}, \{x^{R\mu}_n\})$ of the
collection of modes of the left and the right half of the string. 
(The sliver is an example of such a state). 
We will use vector notation to represent these collections
of modes
\ben    
&&\xl \equiv  
\{ x^{L\mu}_n \, | \, n=1, \cdots \infty; \, \mu = 0, \cdots
25\} \,, \nonumber \\
&&\xr \equiv   
\{ x^{R\mu}_n \, | \, n=1, \cdots \infty; \, \mu = 0, \cdots
25\} \,.
\een
We can also
regard the function
$\psi(\xl, \xr)$ 
as an
infinite dimensional matrix, with the row index labeled by the
modes in 
$\xl$ and the column index labeled by the modes in $\xr$. 
The reality condition on the string
field is the hermiticity of this matrix:
\be \label{eherm}
\psi^*(\xl, \xr)= \psi(\xr, \xl)\, ,
\ee
where the $^*$ as a superscript denotes complex conjugation.
Twist symmetry, on the other hand,
exchanges the left and right
half-strings, so it acts as transposition of the matrix:
twist even (odd) string
fields correspond to symmetric (antisymmetric)
matrices in half-string space.
Furthermore, given two such
functions
$\psi(\xl, \xr)$ and $\chi(\xl, \xr)$, their
$*$ product is given by \cite{WITTENBSFT} 
\be \label{eg9}
(\psi*\chi) \,(\xl, \xr) = 
\int 
[\hbox{d} y]\,\,
\psi(\xl,  y) \,\chi ( y, \xr)\, .
\ee
Thus in this notation
the $*$-product becomes a generalized 
 matrix product. It is clear
that the vector space on which these matrices act is the space
of functionals of the half-string coordinates $\xl$ (or
$\xr$). A projection operator $P$ into a one dimensional subspace of
the half string Hilbert space, spanned by 
some appropriately normalized
functional $f$, will correspond to a functional of the form:
\be \label{eg10}
\psi_P(\xl,
\xr) = f(\xl) f^*(\xr)\, .
\ee
The two factors in this expression are related by conjugation
in order to satisfy condition \refb{eherm}.
The condition $\psi_P * \psi_P = \psi_P$ requires that
\be
\label{normc}
\int [\hbox{d}y]  f^*(y)  f(y) = 1 \,.  
\ee
By the formal properties of the original 
open string field theory 
construction
one has  $\langle A, B\rangle = \int  A*B$ where $\int $ has the
interpretation of a trace, namely identification of the left
and right halves of the string,
together 
with an integration over 
the string-midpoint coordinate $x^{M \, \mu}=X^\mu(\pi/2)$. 
Applying this 
to a projector
$P$ with associated wavefunction $\psi_P (x^L, x^R)
= f^*(x^L) f(x^R)$, and focusing only in the matter sector
    we would find 
\ben
   \langle P, P \rangle &=& \int P*P = \int P = V 
\int [\hbox{d}x^L][\hbox{d}x^R] \delta (x^L
- x^R)
    \psi_P (x^L, x^R) 
\\   
& =& V\int [\hbox{d}y] f^*(y) f(y)   
= V\, ,
   \nonumber  \een
where $V$ is the space-time volume coming from integration over 
the string midpoint $x^{M \, \mu}$. 
   This shows that (formally) rank-one 
projectors are expected to
    have BPZ normalization $V$. 
In our case, due to conformal anomalies,
    while the matter sliver squares precisely as a projector, its BPZ
    norm approaches zero as the level is increased~\cite{0102112}. 
The above argument applies
    to string fields at zero momentum, thus the alternate projector
    constructed 
in ref.\cite{0102112} representing lower dimensional D-branes
need not have the same BPZ norm as the sliver.

\subsubsection{The left-right factorization 
of the sliver wavefunctional} \label{smds2}

The sliver is a projector operator in the space of 
half-string functionals if it factorizes. This 
factorization appears to be indeed true, and can 
be tested numerically\footnote{The factorization
of a related sliver functional suitable for D-instantons
has been proven in \cite{0105059}.}. For this we need to express the
sliver wave-function as a function of
$x^{L\mu}_n$,
$x^{R\mu}_n$ and then see if it
factorizes in the sense of eq.\refb{eg10}. 
For this purpose, we need the position eigenstate
\be \label{eg13}
\langle 
\vec x | = K_0^{26} \, \langle 0|
\exp\Bigl(-x \cdot E^{-2} \cdot x 
+  2 i \,a\cdot E^{-1} \cdot x + {1\over 2} a\cdot
a\Bigr)\, ,\quad E_{nm} = \delta _{nm} \sqrt{{2\over n}} \,.
\ee
The sliver wave-function is then found to be \cite{0105058}
\be \label{eg14}
\psi_\Xi(\vec x 
) = \langle \vec x \, 
| \Xi\rangle
= \wt\NN^{26} \exp \Bigl(-{1\over 2}\, x\cdot 
V \cdot x \Bigr)\, ,\quad
 V_{mn} = n
\delta_{mn} - 2\sqrt{mn} (S(1+S)^{-1})_{mn}\, .
\ee
We can now rewrite $\psi_\Xi$ as a function of $x^L$ and
$x^R$ using eq.\refb{eg5}. This gives:
\be \label{eg16}
\psi_\Xi (\xl, \xr) = \wt\NN^{26} 
\exp\Bigl(-{1\over 2}\xl\cdot K\cdot \xl -{1\over 2} \xr\cdot K \cdot
\xr 
- \xl\cdot L\cdot\xr  \Bigr)\, , 
\ee
where
\be \label{eg17}
K = A^{+T} V A^+ = A^{-T} V A^-, \qquad L = A^{+T} V A^-\, .
\ee
The equality of the two forms for $K$ follows
from eq.\refb{egtwist} and the relation $CSC=S$. The superscript
$T$ denotes transposition.
The sliver wave-function
factorizes if the matrix $L$ vanishes. We have
checked using level truncation that this 
indeed appears to be the case. 
In particular, as the level is increased, 
the elements of the matrix $L$
become much smaller than typical elements of the matrix 
$K$.  If $L$ vanishes,
then the sliver indeed has the form given in eq.\refb{eg10} with
\be \label{egfactor}
f(\xl)= \wt\NN^{13} 
\exp(-{1\over 2}\, 
\xl \cdot K \cdot\xl) \, .
\ee
In this form we also see that the functional $f$ is actually real.
This is expected since the sliver is twist even,
and it must then correspond to a
symmetric matrix in half-string space.

\subsubsection{Building orthogonal projectors} \label{smds22}

\smallskip
Given that the sliver describes a projection operator into a one
dimensional subspace, the following question arises naturally :
is it possible to construct a projection 
operator into an orthogonal one
dimensional subspace? If we can construct such a projection operator
$\chi$, then we shall have 
\be \label{eg18}
\Xi * \chi = \chi * \Xi = 0, \qquad \chi * \chi = \chi\, ,
\ee
and $\Xi + \chi$ will satisfy the equation of motion \refb{eg1} and
represent a configuration of two D-25-branes.

{}From eq.\refb{eg10} it is clear how to construct such an orthogonal
projection operator. We simply need to choose a
function $g$ 
satisfying the same normalization condition 
\refb{normc} as
$f$  
and orthonormal to $f$: 
\be \label{eg19}
 \int [\hbox{d}y] \, f^*(y)
g(y) = 0\,,  \quad
 \int 
[\hbox{d}y]\, g^*(y) g(y) = \int 
[\hbox{d}y] \, f^*(y)f(y) \,  = 1 \,, 
\ee
and then define
\be \label{egchi}
\psi_\chi(\xl,
\xr) = g(\xl) g^*(\xr)\, .  
\ee
There are many ways to construct such a function $g$, 
but one simple class of
such functions is obtained by choosing:
\be \label{eg20}
g(x^L) = \lambda_{\mu n} x^{L\mu}_n \, f(x^L)\, \equiv 
\lambda\cdot
x^L \, f(\xl)\,,
\ee
where $\lambda$ is a constant vector.
Since $f(-x)= f(x)$, the function $g(x)$  is orthogonal
to $f(x)$. For convenience we shall choose $\lambda$
to be real. Making use of \refb{egfactor} the normalization condition
\refb{eg19} for $g$ requires:
\be \label{eg21}
{1\over 2}  \lambda \cdot K^{-1} \cdot  \lambda
= 1\, .  
\ee
Additional orthogonal projectors are readily obtained.
Given another function
$h(x^L)$ of the form
\be \label{eg22}
h(x^L) = \lambda'\cdot x^L \, f(x^L)\, ,
\ee
with real $\lambda'$, 
we find another projector orthogonal to the sliver and to
$\chi$ if
\be \label{eg23}  
 \lambda\cdot K^{-1} \cdot \lambda' = 0, \qquad
{1\over 2} \,\lambda'\cdot K^{-1}  \lambda' = 1. \qquad
\ee
Since we can choose infinite set of mutually 
orthonormal vectors of this
kind, we can construct infinite number of projection operators into
mutually orthogonal subspaces, 
each of dimension one. By superposing $N$
of these projection operators 
we get a solution describing $N$ D-branes.

It is instructive to re-express the string state $\chi$, given by
eqs.\refb{egchi} and \refb{eg20} in the harmonic oscillator basis. 
We have, 
\be \label{eg24}  
|\chi\rangle = (\lambda \cdot \hat x^{L})\, (\lambda\cdot
\hat x^{R})
|\Xi\rangle\, .
\ee
Using eqs.\refb{eg6} and the relation
$\hat x = {i\over 2} \, E \cdot ( a -
a^\dagger )$ we rewrite this as
\be \label{eg25}  
|\chi\rangle = -{1\over 4}\, ( \lambda\cdot  \wt A^+ E  
(a - a^{\dagger}) )\, (\lambda \cdot \wt A^- E  
(a -  a^{\dagger})) |\Xi\rangle\, .
\ee
To express this in terms of  creation operators only
we use  $a \,|\Xi\rangle = - S a^{\dagger}\,|\Xi\rangle$ and find
\be \label{eg27}
|\chi\rangle = \Big( - \xi\cdot a^{\dagger} \,\, \wt\xi
\cdot a^{\dagger} + \kappa\Big) |\Xi\rangle\, ,
\ee
where $\wt\xi_{\mu n} = 
\xi_{\mu m}C_{mn} $, and  
\ben \label{eg28} 
\xi = {1\over 2}\, \lambda \cdot \wt A^+ E 
(1 + S)\, ,  \quad
\kappa = 
\xi \cdot (1+S)^{-1} \cdot
\wt\xi\, . 
\een

\newcommand{\xm}{x^M}

\subsection{Multiple D-brane solutions $-$ Algebraic approach}
\label{smds}

Since the operators $\wt A^+$, $\wt A^-$ and $S$ are known explicitly,
eqs.\refb{eg27}, \refb{eg28} give an explicit expression for a string
state which
squares to itself and whose $*$-product with the sliver vanishes.
Since  the treatment of star products as delta functionals that glue
half strings in path integrals could conceivably be somewhat formal,
and also the  
demonstration that the sliver wave-functional factorizes was based on
numerical study,
it is worth  examining the problem algebraically using
the oscillator representation of star products. 
One can give an explicit construction of the state
$|\chi\rangle$ without any reference to the matrices $\wt A^+, \wt
A^-$. For this we take a trial state of the same form as in  
eq.~\refb{eg27}:
\be \label{eg29}
|\chi\rangle = \Big( - \xi\cdot a^{\dagger} \,\,\wt\xi\cdot
a^{\dagger} + \kappa\Big) |\Xi\rangle\, .
\ee
Here  $\xi$ is taken to be an arbitrary
vector to be determined,  $\wt\xi \equiv C \xi$, and 
$\kappa$ is a constant to be determined. 
We shall actually constrain $\xi$ to satisfy
\be \label{eg30}
\rho_1 \xi = 0, \qquad \rho_2 \xi =\xi \, ,
\ee
where the $\rho_i$ are the projector operators
defined in \refb{proj}.\footnote{We believe, supported
by numerical evidence, that $\xi$, as defined
in eq.\refb{eg28}, automatically satisfies eq.\refb{eg30} 
for any $\lambda$. If so, all the  results that follow 
are consistent with those in the previous subsection.}  
The detailed verification that $\chi$ is a projector orthogonal
to $\Xi$ was given in  
\cite{0105058} and had three parts, which we now summarize:

\smallskip
\noindent
(1) We require $\chi*\Xi =0$ and use this to fix $\kappa$. This
gave
\be \label{eg33}
\kappa 
= -\xi^T T 
(1-T^2)^{-1} \xi\, . 
\ee

\noindent
(2) One verifies that $\chi * \chi = \chi$, if $\xi$ is
normalized as
\be \label{eg37}
\xi^T \, (1 - T^2)^{-1} \, \wt\xi = 1\, .
\ee

\noindent
(3) One confirms that $\langle \chi | \chi \rangle = \langle \Xi
| \Xi \rangle$ and that $\langle \chi | \Xi\rangle=0$. 

\bigskip

The above results imply that the solution described by $\chi$ 
has the same tension as the solution
described by $\Xi$. Since $
\langle \Xi | \chi\rangle = 0$, 
the BPZ norm of $|\Xi\rangle + |\chi\rangle$ is $2\langle\Xi |
\Xi\rangle$. This shows that $|\Xi\rangle + |\chi\rangle$ represents a
configuration with twice the tension of a single 
D-25-brane.\footnote{The new projector 
$\chi$ can be shown to be related to
$\Xi$ by a  rotation in the $*$-algebra.}

\medskip
Consider now another projector $\chi'$ built just as $\chi$ but using
a vector $\xi'$:
\be \label{egp29}
|\chi'\rangle = \Big( - \xi'\cdot a^\dagger \, 
\wt\xi'\cdot a^\dagger + \kappa'\Big) |\Xi\rangle\, ,
\quad\rho_1 \xi' = 0, \quad \rho_2 \xi' =\xi' \, ,
\ee
and 
\be \label{egpaa1}
\kappa' = -\xi^{\prime T} T (1-T^2)^{-1} \xi'\, ,\quad
\xi^{\prime T} (1 - T^2)^{-1}\xi' = 1\, .
\ee
Thus $\chi'$ is a projector 
orthogonal to $\Xi$. In addition, $\chi'$ projects
into a subspace orthogonal to
$\chi$
if  $\chi * \chi'$ vanishes. 
A short computation shows that this requires:
\be \label{egpn1}
\xi^{T} (1-T^2)^{-1} \xi' = 0\, .
\ee
The above equation being  
symmetric in $\xi$ and $\xi'$, it is clear that
$\chi'*\chi$ also vanishes when eq.\refb{egpn1} is satisfied.
We also have: 
\be \label{egpn2}
\langle\chi'|\chi'\rangle = 1, \quad \langle\chi|\chi'\rangle=
\langle\Xi|\chi'\rangle = 0\, .
\ee
Thus $|\Xi\rangle+|\chi\rangle+|\chi'\rangle$ 
describes a solution with
three D-25-branes. 
This procedure can be continued 
indefinitely to generate solutions with
arbitrary number of D-25-branes.

\sectiono{D-$p$ branes in flat space in the algebraic approach}
\label{smds4}

In the previous 
subsections we have described algebraic methods for
constructing
space-time independent solutions of the 
matter part of the field equation
$\Psi * \Psi = \Psi$
which have vanishing $*$-product with the sliver
and with each other.  
Taking the superposition
of such 
solutions and the sliver we get a solution representing 
multiple 
D-branes. In this subsection 
we shall discuss similar construction for
the D-branes of lower dimension.

Explicit solutions 
of the field 
equations representing D-$p$-branes of
all $p\le 25$ have been given in ref.\cite{0102112}. 
 Thus,
for example, if $x^\bmu$
denote directions tangential to the D-$p$-brane 
($0\le \bar\mu\le p$) and  
$x^\alpha$ denote
directions transverse to the D-brane ($p+1\le \alpha\le 25$), 
then a 
solution representing the
D-$p$-brane has the form\cite{0102112}:
\be \label{esq1}
|\Xi_p\rangle = \NN^{p+1} \exp\Big(-{1\over 2}\eta_{\bmu\bnu} S_{mn}
a_m^{\bmu\dagger} a_n^{\bnu\dagger}\Big)|0\rangle
\otimes (\NN')^{25-p} \exp\Big(-{1\over 2} S'_{mn}
a_m^{\alpha\dagger} a_n^{\alpha\dagger}\Big)|\Omega\rangle\, ,
\ee
where in the second exponential the sums 
over $m$ and $n$ run from 0 to
$\infty$,
$\NN'$
is an appropriate normalization constant 
determined in ref.\cite{0102112}, $a_0^\alpha$,
$a_0^{\alpha\dagger}$ are
appropriate linear combinations of the center of momentum
coordinate 
$\hat x^\alpha$ and its conjugate momentum 
$\hat p^\alpha$ 
satisfying
commutation relations of creation and annihilation operators,
and $S'$ is given by an
equation identical to the one for $S$ 
(see  eqs.\refb{ea1}-\refb{ea44a})  
with all matrices $M^{rs}$, $V^{rs}$, $X$, $C$ and $T$ replaced by the
corresponding primed matrices. 
The primed matrices carry indices running
from 0 to $\infty$ in contrast with the unprimed matrices whose indices
run from 1 to $\infty$. But otherwise the primed matrices satisfy the
same equations as the unprimed matrices. Indeed, all the equations in
section \ref{s3.2} are valid with unprimed matrices
replaced by primed matrices, $|\Xi\rangle$
replaced by $|\Xi_p\rangle$ and
$\beta\cdot a^\dagger$ interpreted as
$(\beta_{n\bmu}a^{\bmu\dagger}_n+ 
\beta'_{n\alpha}a^{\alpha\dagger}_n)$. In
particular we can define 
a pair of projectors 
$\rho_1'$ and $\rho_2'$ 
in a manner analogous to
eq.\refb{proj}. 
We now choose 
vectors $\xi_{\bmu m}$, $\xi'_{\alpha m}$
such
that
$\rho_1 \xi_{\bmu} = 0$ and $\rho_1' \xi'_{\alpha} = 0$.
We also define
\be \label{esq3}
\wt \xi_\bmu = C  \xi_\bmu, \qquad \wt \xi'_\alpha = C' \xi'_\alpha,
\qquad
\kappa' = -\xi^T T (1-T^2)^{-1} \xi - \xi^{\prime T} T'
(1-T^{\prime 2})^{-1} \xi'\, ,
\ee
and normalize $\xi$, $\xi'$ such that
\be \label{esq4}
\xi^T (1 - T^2)^{-1} \xi + \xi^{\prime T} 
(1-T^{\prime 2})^{-1} \xi' = 1\, .
\ee
In that case following the procedure used in
subsection \ref{smds} 
we can show that the state:
\be \label{esq5}
|\chi_p\rangle = \Big( - (\xi_{\bmu } \cdot a^{\bmu\dagger} +
\xi'_{\alpha }\cdot
a^{\alpha\dagger}) ( \wt\xi_{\bnu }\cdot
a^{\bnu\dagger} + \wt\xi'_{\alpha }\cdot a^{\alpha\dagger}) +
\kappa'\Big)
|\Xi_p\rangle\, ,
\ee
satisfies:  
\be \label{esq6}
\chi_p * \Xi_p = \Xi_p * \chi_p = 0\, ,\quad\hbox{and}\quad
\chi_p * \chi_p =
\chi_p\, .
\ee
Thus $\chi_p+\Xi_p$ will describe a 
configuration with two D-$p$-branes.
This construction can  be generalized easily  
following the
procedure of subsection \ref{smds} 
to multiple D-$p$-brane solutions.

This procedure can be generalized to construct superposition of
parallel separated D-branes as well as D-branes of different
dimensions\cite{0105058}. But since the general construction for
superposition of D-branes described by
arbitrary boundary conformal field theories has been discussed in
section \ref{sbound}, we shall not give this algebraic
construction
here.

\sectiono{Outlook}

We now offer some brief remarks on our results and discuss some
of the open questions.

\begin{itemize}

\item
Conventional OSFT requires a choice of background
affecting the form of the quadratic term in the action. In that sense
the VSFT action \refb{eo1} represents the choice of the
tachyon vacuum as the background around which
we expand. But 
this clearly is a
special background being the end-point of tachyon
condensation of any D-brane. The VSFT action is formally 
independent of the choice of BCFT used to expand the
string field since $\QQ$ is made purely of ghost operators, and the
$*$-product, defined through overlap conditions on string
wave-functionals, is formally independent of the choice of open
string background. For backgrounds related by exactly
marginal deformations, this notion of manifest background independence
can be made precise using the language of connections in theory space 
\cite{RSZ} 
as has been demonstrated in ref.\cite{0105168}.  
The closed string background dependence of VSFT
deserves investigation and may illuminate the way
closed string physics should be incorporated
\cite{Zwiebach:1998fe,Strom}.

\item The structure of the string field theory action \refb{eo1}
is very similar in spirit to the action of $p$-adic string
theory~\cite{padic, joe}.  
Both are non-local, and 
in both cases the action expanded
around the tachyon vacuum is perfectly non-singular and has no
physical excitations. Yet in both cases the 
theory admits lump solutions
which support open string excitations. The D-$p$-brane solutions are
gaussian in the
case of $p$-adic string theory, and also in the case of VSFT, 
although in this case the string field has additional
higher level excitations. The similarities may extend to
the quantum level, as discussed recently by Minahan~\cite{joe}.

\item 
Vacuum string field theory
is much simpler than conventional cubic open string field theory.
Explicit analytic solutions of 
equations of motion are possible.
Also in this theory off-shell `tachyon'
amplitudes (and perhaps other amplitudes as well) around the tachyon
vacuum can be computed exactly up to overall
normalization. 
This indicates that we are indeed 
expanding the action around a simpler background. 
Even the $p$-adic string action 
takes a simple form only
when expanded about the tachyon vacuum.

\item
The vacuum SFT incorporates
nicely the most attractive features of boundary 
SFT $-$ the automatic generation of correct tensions, 
and the description of solutions in terms of renormalization
group ideas. These features arise in vacuum string field
theory by taking into account the unusual geometrical definition
of the sliver state.
As in boundary SFT, two dimensional field theories with
non-conformal boundary interactions play a role.
However, rather than using them to define the configuration space
of string fields, we use them to construct solutions of the
equations of motion, as 
reviewed briefly in section
\ref{newsection}.

\item
Our work also
gives credence to the idea that half-string functionals do play
a role in open string theory. At least for 
zero momentum string fields,
as explained here, it is on the space of half-string functionals
that the sliver is a rank-one projection operator. 
We have also  
learned how to construct higher rank projectors.
This allows us to construct explicit solutions representing
superpositions of D-branes of various dimensions in the
oscillator representation.

\item 
The identity string
field, on the other hand is an infinite rank projector. 
Since rank-$N$
projectors are associated to configurations with $N$ D-branes, 
one would be led to believe that the identity 
string field is a classical
solution of VSFT representing a background with
an infinite number of D-branes. While technical complications
might be encountered in  
discussing concretely such background, it
is interesting to note that a background with infinite number of
D-branes 
is natural for a general  K-theory analysis of 
D-brane states \cite{0007175}.

\item In the study of $C^*$ algebras and von Neumann algebras
projectors play a central role in elucidating their structure.
Having finally found how to construct (some) projectors in the
star algebra of open strings, a more concrete understanding of
the gauge algebra of open string theory, perhaps based on
operator algebras\footnote{For readable introductory comments on
the possible uses of $C^*$ algebras in $K$-theory see
\cite{0102076}, section 4.}, may be possible to attain in the
near future. This would be expected to have significant impact on
our thinking about string theory.

\item 
Clearly the most
pressing problem at this stage is understanding the ghost sector. This
is needed not merely to complete the construction of the action, but
also  for
understanding gauge transformations in this theory. This, in turn is
needed  for classifying inequivalent classical 
solutions and the spectrum
of physical states around D-brane backgrounds. 
The knowledge of $\QQ$ will
also enable us to calculate quantum effects in 
this theory and determine
whether the theory contains in its full spectrum, the 
elusive closed string states. 

\item Another challenging problem at this stage is to understand
which of the various solutions are gauge equivalent. Some progress
to this direction have been made in ref.\cite{0105168}, but a full
understanding is lacking. Similar problems in the context of
non-commutative field theories have been discussed in
ref.\cite{0105242}.

\end{itemize}

\medskip
All in all
we are in the surprising position where we realize that
in string field theory some non-perturbative physics -- such as 
that related to multiple D-brane configurations -- could be argued
to  emerge more simply than the 
analogous phenomena does in ordinary field
theory.
Thus we are led to believe that vacuum string field theory may
provide a
surprisingly powerful and flexible 
approach to non-perturbative string
theory.

\medskip
\bigskip

\noindent{\bf Acknowledgements}:  
We would like to thank J. David,
D. Gaiotto, R. Gopakumar,  F. Larsen, 
J. Minahan, S. Minwalla, N. Moeller,
P. Mukhopadhyay, M.~Schnabl,
S. Shatashvili, S. Shenker, 
 A. Strominger,
W.~Taylor, E.~Verlinde and E. Witten for useful discussions.
The work of L.R. was supported in
part by Princeton University
``Dicke Fellowship'' and by NSF grant 9802484.
The work of A.S. was supported in part by NSF grant PHY99-07949.
The work of  B.Z. was supported in part
by DOE contract \#DE-FC02-94ER40818.
Finally we thank the organizers  
of Strings 2001 for organizing an
excellent conference.

\end{document}